\documentclass[12pt]{amsart}

\newcommand{\CC}{\mathcal C}
\newcommand{\DD}{\mathcal D}
\newcommand{\pp}{{\bf p}}
\newcommand{\xx}{{\bf x}}

\begin{document}
\title[Pseudodynamical evolution and path integral]{Pseudodynamical evolution and path integral
in quantum field theory}
\author{A. V. Stoyanovsky}
\address{Russian State University of Humanities}
\email{stoyan@mccme.ru}
\begin{abstract}
Using the notion of distribution on an infinite dimensional space defined in
our previous paper, we give definition of a version of dynamical evolution in
quantum field theory, motivated by heuristic formulas involving path integrals.
\end{abstract}
\maketitle

\section*{Introduction}

The problem of defining dynamical evolution in quantum field theory (QFT) is rather old.
Usual approaches to it using the half $S$-matrix meet difficulties such as
non-renormalizable
multiloop divergencies in perturbation theory. In this note we present one more
approach to dynamical evolution in QFT based on the previously defined notion of
distribution on a space of functions [1]. Our approach is motivated by a heuristic
argument with path integrals. Note that the notion of distribution on
a space of functions somehow formalizes the notion of path integral, but
this formalization seems not sufficient for the argument presented in this note.
This argument relates dynamical evolution with the generating functional of Green
functions via Fourier transform.

The author is obliged to V. P. Maslov for helpful discussions.

\section{Representation of dynamical evolution via path integral}

Consider first a (linear) partial differential equation of the form
\begin{equation}
\frac{\partial\psi}{\partial t}=A\psi,
\end{equation}
where $\psi=\psi(t,q)$ is an unknown (possibly vector) function, $q=(q_1,\ldots,q_n)$,
and the right hand side $A\psi$ is a partial differential expression.
Then, an old and well known
problem is to represent the dependence of solutions $\psi(T,q)$
on the initial data, say $\psi(T_0,q)$, by a path integral:
\begin{equation}
\psi(T,q)=\int\limits_{q_0}
\int\limits_{\begin{subarray}{c}q(t),\ T_0\le t\le T\\q(T_0)=q_0,q(T)=q\end{subarray}}
D\mu_1^{q_0,q}(q(\cdot))\ \psi(T_0,q_0)dq_0.
\end{equation}
Here $D\mu_1^{q_0,q}(q(\cdot))$ is a ``measure'' on the space of trajectories $q(t)$,
$T_0\le t\le T$,
starting at the point $q_0$ and ending at the point $q$.

Let us formally manipulate with expression (2) a little as if it were
mathematically well defined. Let us make Fourier transform of (2) with respect to $q(t)$
and with respect to the endpoints $q_0$ and $q$, i.~e., let us go to ``$p$-representation with
external force $j_1(t)$''. This means, in particular, that we consider more general
equation
\begin{equation}
\frac{\partial\psi}{\partial t}=A\psi+ij_1(t)q\psi
\end{equation}
(with $i=\sqrt{-1}$) instead of (1).

We obtain
\begin{equation}
\begin{aligned}{}
&\hat \psi(T,p)\stackrel{\text{def}}=\int\limits_q \psi(T,q)e^{ipq}dq\\
&=\int\limits_q\int\limits_{q_0}
\int\limits_{\begin{subarray}{c}q(t),\ T_0\le t\le T\\q(T_0)=q_0,q(T)=q\end{subarray}}
D\mu_1^{q_0,q}(q(\cdot))e^{i\left(\int_{T_0}^Tj_1(t)q(t)dt+pq\right)}\psi(T_0,q_0)dq_0dq\\
&=\frac1{(2\pi)^n}\int\limits_q\int\limits_{q_0}
\int\limits_{\begin{subarray}{c}q(t),\ T_0\le t\le T\\q(T_0)=q_0,q(T)=q\end{subarray}}
\int\limits_{p_0}D\mu_1^{q_0,q}(q(\cdot))dqdq_0e^{i\left(\int_{T_0}^T
j_1(t)q(t)dt+pq-p_0q_0\right)}\\
&\times\hat \psi(T_0,p_0)dp_0,
\end{aligned}
\end{equation}
hence
\begin{equation}
F_{q\to p}F_{q_0\to p_0}U(\mu_L^{q_0}\mu_R^{q}F_{q(\cdot)}\mu_1^{q_0,q}(j_1(\cdot)))
=\frac1{(2\pi)^n}U(F\mu(j(\cdot))),
\end{equation}
where $\mu$ is the ``measure'' on the space of unbounded trajectories $q(t)$,
$-\infty<t<\infty$, $F\mu$ is the Fourier transform of the measure $\mu$,
which is a functional of a (vector valued) function $j(t)$, $-\infty<t<\infty$,
and $j(t)$ is the following (vector valued) distribution:
\begin{equation}
j(t)=p\delta(t-T)-p_0\delta(t-T_0)
+\left\{\begin{array}{l}j_1(t),\ \ T_0\le t\le T,\\
0,\ \ \text{otherwise.}\end{array}\right.
\end{equation}
Further, $\mu_L^{q_0}$ is the ``measure'' of the space of infinite trajectories $q(t)$,
$-\infty<t\le T_0$, bounded from the right and ending at the point $q_0$, $q(T_0)=q_0$,
and $\mu_R^q$ is the ``measure'' of the space of infinite trajectories $q(t)$,
$T\le t<\infty$, bounded from the left and starting at the point $q$, $q(T)=q$.
The symbol $U$ means the integral operator with the kernel in brackets depending on
$q_0,q$ or on $p_0,p$. $F$ always means Fourier transform with respect to the shown arguments.

Thus,
we have formally reduced the problem of presenting dynamical evolution in the form of
a path integral to the problem of defining path ``measure'' for unbounded trajectories and its
Fourier transform.

A trivial, in a sense, solution to this latter problem has been given in [1].
According to {\it loc.~cit.},
Fourier transform establishes an equivalence between 1)~``measures'',
or, more precisely, {\it distributions} $\mu$ on appropriate linear space of functions,
and 2)~certain functionals, or, more precisely, formal power series $F\mu$
on the dual linear space. This $F\mu$ coincides with the {\it generating functional of
Green functions} of the distribution.

Since the ``measure'' $\mu_1^{q_0,q}$ ``tends'' to
$\mu$ as $T_0\to-\infty$, $T\to\infty$, one can assume that
the ``measure'' $\mu$ is determined at most uniquely
by the ``measure'' $\mu_1^{q_0,q}$. Hence, one can assume that
the Green functions of equation (1) are determined
at most uniquely by this equation. However, it remains unclear how to
find all equations (1) which admit Green functions in the sense above,
and how to find these Green functions for these equations, and whether one can find
them at all. In this sense, it remains unclear whether quantum field theory, in which
Green functions are well defined, is a direct generalization of the theory of
linear partial differential equations to the case of multidimensional trajectories.

The problem of presentation of solutions of equation (1) in the form of a path integral could
be solved directly, for example, in the $p$-representation,
as it is done in [2] for the Schrodinger equation.

\section{Application to quantum field theory}

Assume now that we are considering a model of quantum field theory (QFT) in a space-time
with coordinates $x=(x_0,\ldots,x_n)$, $x_0=t$. It is natural to
assume that the dynamics should be formally given by formula analogous to (2) in the space
of distributions [1] $\Psi(\CC,q)$ on the space of tempered distributions $q=q(s)$ on a
space-like surface $\CC$ given by smooth parameterization $x=x(s)$, $s=(s_1,\ldots,s_n)$:
\begin{equation}
\Psi(\CC,q)=\int\limits_{q_0}
\int\limits_{\begin{subarray}{c}v(x),\ x\in\DD\\v|_{\CC_0}=q_0,v|_\CC=q\end{subarray}}
D\mu_1^{q_0,q}(v(\cdot))\ \Psi(\CC_0,q_0)Dq_0,
\end{equation}
where $\DD$ is the region between the two space-like surfaces $\CC_0$ and $\CC$.

Introducing the source $j_1(x)$ and calculating exactly as in (4,5),
we obtain the following formal relation:
\begin{equation}
F_{q\to u}F_{q_0\to u_0}U(\mu_L^{q_0}\mu_R^{q}F_{v(\cdot)}\mu_1^{q_0,q}(j_1(\cdot)))
=\frac1{(2\pi)^\infty}U(F\mu(j(\cdot))),
\end{equation}
where $F\mu(j(\cdot))$ is the generating functional of the Green functions of the given model
of QFT, and $j(\cdot)$ is the following distribution:
\begin{equation}
j(x)=u\delta_\CC(x)-u_0\delta_{\CC_0}(x)
+\left\{\begin{array}{l}j_1(x),\ \ x\in\DD,\\
0,\ \ \text{otherwise.}\end{array}\right.
\end{equation}
Here $\delta_\CC(x)$ is the delta function of the surface $\CC$.
Further, the symbols $\mu_L^{q_0}$, $\mu_R^q$, $U$ have the sense similar to that of (5).

The formulas above are heuristic and formal. However, we can deduce from them the following
principle.
\medskip

{\bf Principle.} {\it If we substitute into the generating functional of the Green functions
of a model of QFT the source $j(x)$ of the form
\begin{equation}
j(x)=u\delta_\CC(x)-u_0\delta_{\CC_0}(x),
\end{equation}
for two space-like surfaces $\CC$ and $\CC_0$, then, as the surface $\CC$ varies,
the resulting functional of a function $u=u(s)$ on the surface $\CC$ represents the
Fourier transform of certain version of dynamical evolution of the given model of QFT.}
\medskip

This principle can be considered as a first version of {\it definition} of dynamical
evolution in quantum field theory. We call this evolution by {\it pseudodynamical evolution}.
It is obtained from the ``true'' dynamical evolution by ``multiplying the wave functionals
$\Psi(\CC_0,q_0)$, $\Psi(\CC,q)$ by the factors $\mu_L^{q_0}$, $\mu_R^q$ respectively''.
\medskip

{\it Example: free scalar field.} Consider the generating functional of the Green functions
of free scalar field:
\begin{equation}
F\mu(j)=\exp\frac{-i}{2h}\int\frac{\hat j(p)\hat j(-p)}{p^2-m^2+i\varepsilon}dp,
\end{equation}
where $p=(p_0,\ldots,p_n)$, $p_0=E$, $p^2=E^2-\pp^2$, $\pp=(p_1,\ldots,p_n)$.

Let us substitute into it the source of the form
\begin{equation}
j(x)=u(\xx)\delta(t-T)-v(\xx)\delta(t),
\end{equation}
where $\xx=(x_1,\ldots,x_n)$. Then a direct calculation shows that the resulting functional
of $T$ and $u(\xx)$, denote it by $\Phi(T,u(\cdot))$,
satisfies, up to multiplication by a function
of $T$, the Fourier transform (with respect to $u(\cdot)$)
of the normally ordered QFT Schrodinger equation:
\begin{equation}
ih\frac{\partial\bar\Phi}{\partial T}=:\int\frac12\left(u(\xx)^2
-h^2\left(\nabla\frac{\delta}{\delta u(\xx)}\right)^2
-h^2m^2\frac{\delta^2}{\delta u(\xx)^2}\right)d\xx\ \bar\Phi :,
\end{equation}
where $\bar\Phi=\Phi\cdot f(T,v(\cdot))$.

This functional also exactly satisfies the following differential equation
(in the $\pp$-representation), with $h=1$:
\begin{equation}
i\frac{\partial\Phi}{\partial T}=\int \hat u(\pp)
\left(\sqrt{\pp^2+m^2}\frac{\delta}{\delta\hat u(\pp)}
-\hat u(-\pp)\right)d\pp\ \Phi.
\end{equation}

\end{document}